\documentclass[pre,twocolumn,a4paper,showpacs]{revtex4}
\usepackage{amssymb}
\usepackage{epsfig}
\begin{document}
\title{Statistics of opinion domains of the majority-vote model on a square lattice}

\author{Lucas R. Peres and Jos\'e F. Fontanari}
\affiliation{Instituto de F\'{\i}sica de S\~ao Carlos,
  Universidade de S\~ao Paulo,
  Caixa Postal 369, 13560-970 S\~ao Carlos, S\~ao Paulo, Brazil}

\begin{abstract}

The existence of juxtaposed regions of  distinct cultures in  spite of the fact that 
people's beliefs have a tendency
to become more similar to each other's as the individuals interact repeatedly is a puzzling 
phenomenon in the social sciences. Here we study an extreme version of
the frequency-dependent bias model of social influence in which an individual adopts the  opinion 
shared by the majority of the members of its extended neighborhood, which includes the individual itself. This is  
a  variant of  the majority-vote model in which the individual retains its opinion in case there is a tie  among the neighbors' opinions.
We assume that the individuals are fixed in the sites of a square
lattice of linear size $L$ and that they  interact with their  nearest neighbors only. 
 Within a mean-field framework, we derive the equations 
of motion for  the density of individuals adopting a particular opinion in the single-site and pair approximations. 
Although the single-site approximation predicts a single opinion domain that takes over the entire lattice, the pair approximation yields 
a qualitatively correct picture with the coexistence of  different opinion domains and a strong dependence on the initial conditions. 
Extensive Monte Carlo simulations indicate the existence of a rich distribution of  opinion domains or clusters, 
the number of which grows with $L^2$  whereas the size of the largest cluster grows with  $\ln L^2$.  
The analysis of the sizes of the opinion domains shows that they obey a power-law distribution for not too large sizes but that they are 
exponentially distributed in the limit of very large clusters. In addition, similarly to other well-known social influence 
model -- Axelrod's model --  we found that these opinion domains are unstable to the effect of a thermal-like noise.

\end{abstract}

\pacs{89.75.Fb, 87.23.Ge, 05.50.+q}
\maketitle

%
\section{Introduction} \label{sec:Intro}

Despite the tendency of social actors to become more similar 
to each other through their local interactions \cite{Latane_81,Moscovici_85}, global polarization, i.e., the existence of
stable multicultural  regimes, is a ubiquitous phenomenon in human society \cite{Boyd_85,Nowak_90}.  
This issue has been addressed by somewhat idealized agent-based simulations of human behavior producing 
highly nontrivial insights on the nature of the collective behavior that  results from the homogenizing local
interactions (see \cite{RMP} for a recent review). 

In this line, a particularly successful model is Axelrod's model for the dissemination of culture or social influence 
\cite{Axelrod_97}.
In Axelrod's model, the agents  are placed at the sites of a square lattice of size $L \times L$ and can interact  with their nearest  neighbors only. 
The culture of an agent 
is represented by a string of $F$ cultural features, where each feature can adopt a certain number $q$ of distinct traits. 
The interaction between any two neighboring agents takes place with probability proportional  to the number of 
traits they have in common. Although the result of such interaction is the increase of the similarity between 
the two agents, as one of them modifies  a previously  distinct trait to match that of its partner, the model exhibits global 
polarization \cite{Axelrod_97}. The key ingredient for the existence of  stable  globally polarized 
states in Axelrod's model is  the rule that prohibits the interaction between  agents which
do not have any  cultural trait in common. Relaxation of this rule so as to permit interactions
regardless of the similarity between agents leads to one of the $q^F$ distinct
absorbing homogeneous configurations \cite{Kennedy_98}.  In addition, introduction of external noise so that
traits can change at random with some small probability  \cite{Klemm_03c}, as well as the increase of the connectivity of the agents
\cite{Greig_02,Klemm_03b,Klemm_03a},  destabilizes the polarized state also.

Although  Axelrod's model enjoyed great popularity among the statistical physics community due mainly to 
the existence of a non-equilibrium phase transition \cite{Marro_99} that separates the globally homogeneous from the globally polarized regimes 
\cite{Castellano_00,Barbosa_09}, the vulnerability   of the  polarized absorbing configurations
was considered a major drawback to explaining  robust collective social behavior. In this vein,
Parisi et al.  \cite{Parisi_03} have proposed a lattice version of the classic frequency bias mechanism   
for cultural or opinion change \cite{Boyd_85,Nowak_90}, which
assumes that  the number of people holding an opinion is the key factor for an agent to adopt that opinion, i.e., people
have a tendency to espouse cultural traits that are more common in their social environment. Actually, almost any model of cultural 
transmission admits that the probability that an individual acquires a cultural variant depends on the frequency of that 
variant in the population. The frequency-dependent bias mechanism requires, however, that the individual be {\it disproportionately}
likely to acquire the more common variant \cite{Boyd_85}.

More to the point, in the model of Parisi et al. 
the culture of an agent is specified by a binary  string of length $F$ (so $q=2$) and each bit of that
string takes the value which is more common among its neighbors  \cite{Parisi_03}. Since the model can be immediately  recognized as $F$ independent majority-vote models  \cite{Liggett_85, Gray_85}, we found the claim by those authors  that such model exhibits a polarized regime most intriguing. 
In order to check whether the multicultural absorbing configurations reported by Parisi et al. were not  artifacts of the small lattice size  
used in their study ($L =20$),  in this paper we present a detailed analysis of the effects of the finite size of the lattice by considering 
square lattices of linear size up to $L = 4000$. In addition, we set $F=1$ to stress the identification with the well-known majority vote 
model of statistical physics. Hence the agents or sites of the lattices are modeled by Ising spins.
 
 Our findings indicate that  the polarized regime 
is indeed  stable in the thermodynamic limit $L \to \infty$ and that the element accountable for this stability is the procedure used to calculate  the majority, which includes the 
site to be updated (target site) in addition to its nearest neighbors. 
For sites in the bulk of the lattice, this procedure  essentially specifies the criterion of update in case of a tie, i.e., 
in case there is no majority among the neighbors. In this case, the majority-vote variant used by  Parisi et al. leaves the state of 
the agent untouched, whereas the most common variant used in the literature sets that state at random with probability $1/2$ \cite{Tome_91,deOliveira_92}. It should be noted, however, that the criterion used by Parisi et al. is actually the original definition of the majority-vote model as introduced in Refs.\ \cite{Liggett_85, Gray_85}. 
It is surprising that such (apparently) minor implementation detail produces nontrivial consequences in the thermodynamic limit, which can actually be confirmed analytically using a mean-field approach that takes into account the correlation between nearest neighbors  -- the pair approximation \cite{Marro_99}.  However, 
the addition of a thermal-like noise such that the individuals may take a stance opposed to the majority opinion with some small probability
destabilizes the globally polarized regime leading to one of the two   low-temperature homogeneous steady states of the  Ising model \cite{Glauber_63}.
In that sense we disagree with the claim by Parisi et al. that their variant of the majority-vote model is more robust than Axelrod's model to the effect of this type of noise.

The rest of this paper is organized as follows. In Sect.~\ref{sec:model} we describe the variant of the majority-vote model proposed by Parisi et al., 
which henceforth we refer to as the extended majority-vote model 
since its key ingredient is  the stretching of the neighborhood to include the target site \cite{Parisi_03}. 
In Sect.~ \ref{sec:MF} we present the mean-field  approach to this model using the single-site and pair approximations
\cite{Tome_91,Marro_99}.
In Sect.~ \ref{sec:res} we use extensive Monte Carlo simulations to investigate several properties of the  absorbing configurations such as 
the average number of opinion domains or clusters, 
the size of the largest cluster and the distribution of cluster sizes, giving emphasis to their dependences on the lattice size. In that section also 
we discuss  the vulnerability of the heterogeneous opinion regime against a thermal-like noise, which allows the individuals to disagree  
with the majority opinion. 
Finally, in Sect.~ \ref{sec:conc} we present our concluding remarks.

\section{Model}\label{sec:model}

The agents are fixed at the sites of a square lattice of size $L \times L$ with open boundary conditions
(i.e., agents at the corners interact with two neighbors, agents at the sides  with three, 
and agents in the bulk with four nearest neighbors). 
The initial configuration is random: the opinion of each agent is set
by a random digit $0$ or $1$ with equal probability. At each time we pick a target  agent at random and then verify
which is the more frequent opinion ($1$ or $0$) among its extended neighborhood, which includes the target agent itself.  The opinion of the target agent is then  changed 
to match the corresponding majority value. We note that there are no ties for agents in the bulk or at the corners of the square
lattice since in these cases the extended neighborhood comprises $5$ and  $3$ sites, respectively. However, agents at the
sides of the lattice have an extended neighborhood of $4$ sites (i.e., $3$ neighbors plus the target agent) and so in case of a tie,
the  opinion of the target agent remains unchanged. Of course, in the limit of large lattices the contribution of these
boundary sites will be negligible. This procedure is repeated until 
the system is frozen in an absorbing configuration.

Although the majority-vote rule or, more generally, the frequency bias mechanism for
opinion change \cite{Boyd_85} is a homogenizing assumption by which the agents become more similar 
to each other, the above-described model does seem to  exhibit global polarization, i.e., heterogeneous absorbing configurations
 \cite{Parisi_03}. Since the study of Ref.\ \cite{Parisi_03} was based on 
a small lattice  of linear size $L=20$ a more careful analysis is necessary to confirm whether this 
conclusion holds in the thermodynamic  limit  as well. It is interesting to note that for 
the more popular variant of the majority-vote model, in which the state of the target site is not included in
the majority reckoning, and  ties are decided by choosing the opinion of the target agent at
random with probability $1/2$,  the only absorbing states in the thermodynamic limit are the two homogeneous configurations. 
For finite lattices, however, we find heterogeneous absorbing configurations  characterized by stripes that
sweep the entire lattice. In Sect.~ \ref{sec:res}  we carry out a finite size scaling analysis of the extended 
majority-vote rule aiming at understanding the nature of the opinion domains   that fragment
the absorbing configurations.

\section{Mean-field analysis}\label{sec:MF}

In this section we offer an analytical approximation to the extended majority-vote model introduced  by Parisi  et al \cite{Parisi_03}. 
The state of the  agent at site  $i$ of the square lattice is 
represented by the binary variable $\eta_i = 0,1$ and so the configuration of the entire lattice comprising $N=L^2$ sites is 
denoted by $\eta \equiv \left ( \eta_1, \eta_2, \ldots, \eta_N \right )$. The master equation that governs the time evolution 
of the probability distribution $P \left ( \eta, t \right )$ is given by
\begin{equation}\label{Master1}
\frac{d}{dt} P \left ( \eta, t \right ) = \sum_i \left [ W_i \left ( \tilde{\eta}^i \right ) P \left (\tilde{\eta}^i, t \right )
- W_i \left ( \eta \right ) P \left (\eta, t \right )  \right ]
\end{equation}
where  $\tilde{\eta}^i = \left ( \eta_1, \ldots, 1- \eta_i, \ldots, \eta_N \right )$ and $W_i \left ( \eta \right )$ is 
the transition rate between configurations $\eta$ and $\tilde{\eta}^i$ \cite{Tome_91,deOliveira_92}.
 For the extended majority-vote model  we have
\begin{equation}\label{W1}
W_i \left ( \eta \right ) = \left  |  \Theta \left [  \sum_\delta \eta_{i+\delta} + \eta_i- 3 \right ]  -
\eta_i  \right |
\end{equation}
where the notation $\sum_\delta \left ( \ldots \right )$ stands for the sum over the 4 nearest neighbors of site $i$ 
and $\Theta \left ( x \right ) =1 $ if $x \geq 0$ and $0$ otherwise.

We are interested in determining the fraction of agents holding opinion $1$ or, equivalently, the fraction  of sites in state $1$, which we denote by $\rho$. Since all sites are 
equivalent we have $\rho \equiv \sum_i \eta_i/N = \left \langle \eta_i \right \rangle $ and this mean value is given by the equation
\begin{equation}\label{R0}
\frac{d}{dt}  \left \langle \eta_i \right \rangle = \left \langle \left ( 1 - 2 \eta_i \right )  W_i \left ( \eta \right )
\right \rangle 
\end{equation}
with $ \left \langle  \left ( \ldots \right ) \right \rangle \equiv \sum_\eta \left ( \ldots \right )  P \left (\eta, t \right ) $ 
as usual. To carry out this average we need to make approximations. In what follows we study in detail two such approximation schemes, 
namely, the single-site approximation  and the pair approximation.

\subsection{The single-site approximation}

This the simplest mean-field scheme which assumes that the sites are independent random variables so that 
Eq. (\ref{R0}) becomes
\begin{eqnarray}\label{R1a}
\frac{d \rho }{dt} &  =  & -\rho  \sum_{n=0}^4 B_n^4 \left ( \rho \right ) \left  |  \Theta \left [  n - 2 \right ]  - 1  \right |
\nonumber \\
&  & + \left ( 1 - \rho \right )  \sum_{n=0}^4 B_n^4 \left ( \rho \right )  \Theta \left [  n - 3 \right ] 
\end{eqnarray}
where $B_n^m$ is the Binomial distribution
\begin{equation}\label{Bin}
 B_n^m \left ( \rho \right )  =  \left ( \begin{array}{c} m \\ n \end{array} \right ) \rho^n \left ( 1- \rho \right )^{m-n} .
\end{equation}
Carrying out the sums explicitly and rearranging the terms yield
\begin{equation}\label{R1b}
\frac{d \rho }{dt} = - \rho \left ( 1 - \rho \right ) \left ( 2 \rho - 1 \right ) \left (  3 \rho^2 - 3 \rho -1  \right ) .
\end{equation}
This equation, which is invariant to the change $\rho \leftrightarrow 1- \rho$, has three fixed points, 
namely, $\rho^* = 0$, $  \rho^* = 1 $ and  $ \rho^* = 1/2$. The first two fixed points are stable and the 
third  one is unstable. This means that in the single-site approximation the only stable configurations are the homogeneous ones.
Finally, we note that $\rho$ contains the same information as the  single-site probability 
distribution $p_1 \left ( \eta_i \right ) $. In fact, $p_1 \left ( \eta_i = 1 \right ) = \rho $ and  $p_1 \left ( \eta_i = 0 \right ) = 1 -\rho $. 

\subsection{The pair approximation}

In this scheme we assume that nearest neighbors sites are  statistically dependent random variables so that, in addition to the 
single-site probability distribution $p_1 \left ( \eta_i \right ) $,  we need to compute the pair probability distribution $
p_2 \left ( \eta_i, \eta_{i+\delta} \right )$ as well. Of particular importance for the subsequent calculations is the conditional 
probability distribution $ p_{1|1} \left ( \eta_{i+\delta} \mid \eta_i \right )$ which is given simply  by the ratio between the pair 
and the single-site probability distributions. 

For the sake of concreteness, let us denote the states of the 4 neighbors of site $i$ by $\eta_1, \eta_2, \eta_3$ and  $\eta_4$. These are 
independent random variables  in the pair approximation since
they are not nearest neighbors in the square lattice, so the sum $n = \eta_1 + \eta_2 + \eta_3 + \eta_4 $ 
that appears in the argument of the Theta functions is a sum of independent variables. With these notations we can rewrite Eq.\ (\ref{R0}) as 
\begin{eqnarray}\label{R2a}
\frac{d \rho }{dt} &  =  & -\rho \sum_{\eta1,\ldots,\eta_4} p_{4 \mid 1} \left ( \eta_1,\eta_2,\eta_3 ,\eta_4 \mid 1 \right ) \left | 
\Theta \left ( n - 2 \right ) -1 \right |  \nonumber \\
&	&  + \left ( 1 - \rho \right ) \sum_{\eta1,\ldots,\eta_4} p_{4 \mid 1} \left ( \eta_1,\eta_2,\eta_3 ,\eta_4 \mid 0 \right )
\Theta \left ( n - 3 \right ) \nonumber \\
\end{eqnarray}
where we have  carried out the sum over $\eta_i =0,1$ explicitly. The assumption that $\eta_1,\dots,\eta_4$ are statistically
independent random variables allows to write
\begin{equation}
p_{4 \mid 1} \left ( \eta_1,\eta_2,\eta_3 ,\eta_4 \mid \eta_i \right ) = p_{1 \mid 1} \left ( \eta_1 \mid  \eta_i  \right ) \times \ldots \times   
p_{1 \mid 1} \left ( \eta_4 \mid  \eta_i  \right )
\end{equation}
and finally obtain
\begin{eqnarray}\label{R2b}
\frac{d \rho }{dt} & =  & -\rho  \sum_{n=0}^4 B_n^4 \left  [ p_{1 \mid 1} \left ( 1 \mid  1  \right ) \right ] \left  |  
\Theta \left [  n - 2 \right ]  - 1  \right | \nonumber \\
&   &  + \left ( 1 - \rho \right )  \sum_{n=0}^4 B_n^4 \left [  p_{1 \mid 1} \left ( 1 \mid  0  \right )  \right ]  \Theta \left [  n - 3 \right ] 
\end{eqnarray}
which is identical to Eq. (\ref{R1a}) except for the arguments of the binomial distributions. Using the notation
$\phi \equiv p_2  \left ( 1,  1  \right )$ we write
\begin{eqnarray}
  p_{1 \mid 1} \left ( 1 \mid  1  \right ) & = &  \frac{\phi}{\rho}  \\
  p_{1 \mid 1} \left ( 1 \mid  0  \right ) & = & \frac{\rho - \phi}{1-\rho}
\end{eqnarray}
so that Eq.\ (\ref{R2b}) involves only two unknowns, $\rho$ and $\phi$. Carrying out the summations explicitly so as to eliminate 
the Theta functions yields
\begin{eqnarray}\label{R2c}
 \frac{d \rho }{dt} & = & -\rho \left [ B_0^4 \left ( \frac{\phi}{\rho} \right ) + B_1^4 \left ( \frac{\phi}{\rho} 
 \right )  \right ]  \nonumber \\
&    & + 
 \left ( 1 - \rho \right ) \left [ B_3^4 \left ( \frac{\rho-\phi}{1-\rho} \right ) + B_4^4 \left ( \frac{\rho - \phi}{1-\rho}   \right )  \right ] .
\end{eqnarray}
We note that this equation reduces to Eq.\ (\ref{R1b}) in the case the neighboring sites are assumed independent, i.e., $\phi = \rho^2$.

\begin{figure}
\centerline{\epsfig{width=0.52\textwidth,file=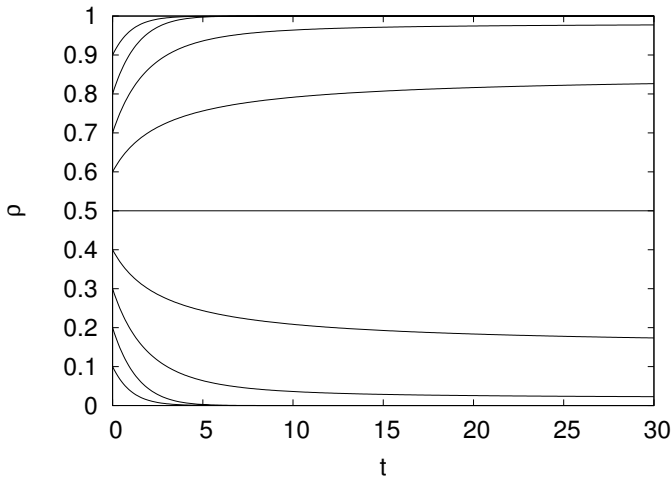}} 
\caption{Time evolution of the fraction of sites in state $1$, $\rho$, in the pair approximation obtained by solving Eqs.\ (\ref{R2c})
and (\ref{R2e}) using Euler's method with step-size $0.01$ for different initial conditions (bottom to top) $\rho_0 = 0.1, \ldots, 0.9$.
\label{fig:1} }
\end{figure}

\begin{figure}[ht]
\centerline{\epsfig{width=0.52\textwidth,file=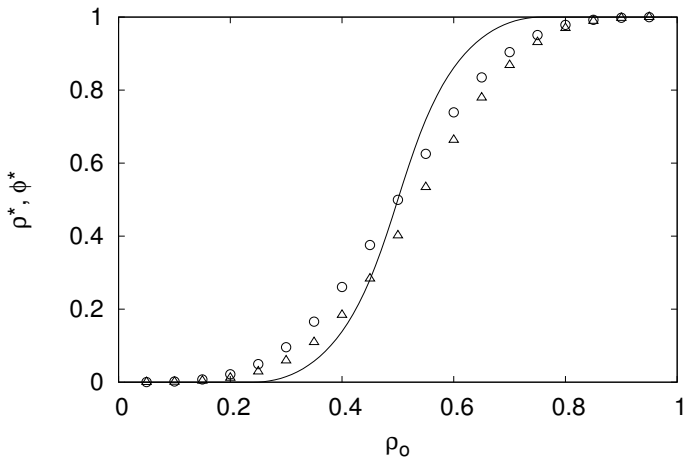}}
\caption{The fraction of sites in state $1$ at equilibrium, $\rho^* $ (represented by $\circ$), and 
the probability that two neighbors are in state $1$ at equilibrium, $\phi^*$ (represented by $\triangle$)
as functions of the initial fraction of sites in state $1$, $\rho_0$. 
The solid line is the result of the pair approximation for which $\phi^* \to \rho^*$.
The initial condition is $\rho = \rho_0$ and $\phi = \rho_0^2$. The symbols
show the results of the Monte Carlo simulations for a square lattice of linear size $L=200$.
\label{fig:2} } 
\end{figure}

Next, our task  is to determine an equation for $\phi =  \left \langle \eta_i \eta_j \right \rangle  $ where $j$ labels one of the four 
nearest neighboring sites of $i$.  We have
\begin{equation}
\frac{d}{dt}  \left \langle \eta_i \eta_j  \right \rangle = 2 \left \langle \eta_j \left ( 1 - 2 \eta_i \right )  W_i \left ( \eta \right )    \right \rangle  .
\end{equation}
Carrying out the average over $\eta_i$ and $\eta_j$ (say $j=1$) explicitly yields
\begin{eqnarray}\label{R2d}
\frac{d \phi }{dt} & =  & -2 \rho p_{1 \mid 1}\left ( 1 \mid  1  \right ) 
\sum_{m=0}^3 B_m^3 \left  [ p_{1 \mid 1} \left ( 1 \mid  1  \right ) \right ] \left  |  \Theta \left [  m - 1 \right ]  - 1  \right | \nonumber \\
&   &  + 2 \left ( 1 - \rho \right )  p_{1 \mid 1} \left ( 1 \mid  0  \right )  
\sum_{m=0}^3 B_m^3 \left [  p_{1 \mid 1} \left ( 1 \mid  0  \right )  \right ]  \Theta \left [  m - 2 \right ] \nonumber \\
\end{eqnarray}
which can be written more compactly as
\begin{eqnarray}\label{R2e}
\frac{d \phi }{dt} &  = & - 2 \phi B_0^3 \left ( \frac{\phi}{\rho} \right ) \nonumber \\
&  & + 2 \left ( \rho - \phi \right ) \left [ B_2^3 \left ( \frac{\rho-\phi}{1-\rho} \right )
+ B_3^3  \left ( \frac{\rho-\phi}{1-\rho} \right ) \right ] . \nonumber \\
\end{eqnarray}
Equations (\ref{R2c}) and (\ref{R2e}) determine completely the time evolution of $\rho$ and $\phi$ and so they are our final equations 
for the pair approximation of the extended majority-vote model.  Figure \ref{fig:1}, which shows
the time evolution of the density of sites in state $1$, confirms  that Eq. (\ref{R2e}) is indeed invariant to the change $ 0 \leftrightarrow 1$.
Most surprisingly, this figure uncovers an unexpected dependence on the choice of the (random) initial condition $\rho_0$ and  $\phi_0 = \rho_0^2$: indeed in the range $\rho_0 \in \left ( \rho_{m},  1 - \rho_m  \right )$,  where $\rho_m \approx 0.25$, the equilibrium 
solution $\rho^*$ is a smooth function  of $\rho_0$, i.e., there is a continuum of fixed points. Accordingly,
Fig. \ \ref{fig:2} shows the fixed point
solutions of Eqs.\ (\ref{R2c}) and (\ref{R2e}) for the usual situation in  which the states of the sites of the initial configuration are set  $1$ with probability $\rho_0$ and
$0$ with probability $1 - \rho_0$. For this setting we have $\phi_0 = \rho_0^2$. 

The mathematical explanation for the odd dynamical behavior exhibited in Figs.  \ref{fig:1} and  \ref{fig:2} is that the dynamic evolution is 
such that $\rho \to \phi$ for $t \to \infty$, a condition that  solves the two equations 
(\ref{R2c}) and (\ref{R2e}) at equilibrium (i.e., $d\rho/dt = d\phi/dt = 0$) simultaneously  
and so leaves one of the unknowns free to take any arbitrary value set by the dynamics or the initial conditions. 
The physical reason is that there are in fact many absorbing configurations very close to the random initial configurations and a few 
lattice updates (typically 10) is sufficient to freeze the dynamics (see Fig.\ \ref{fig:3}). 
In fact, the comparison with the results of the Monte Carlo simulations exhibited in Fig.\ \ref{fig:2} shows a good qualitative agreement between the pair 
approximation and the simulation results. 

 Finally, we note that the pair approximation for the variant of the majority-vote model  in which ties are decided by 
flipping the state of the target site with probability $1/2$ yields the anticipated result that the only stable fixed points are $\rho=\phi = 0$ and $\rho=\phi = 1$, whose basins of attraction depend on whether $\rho_0$ is less or greater than $1/2$.  

\section{Monte Carlo simulations}\label{sec:res}

\begin{figure}
\centerline{\epsfig{width=0.52\textwidth,file=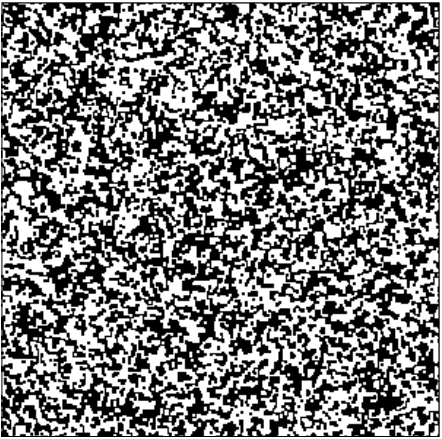}}
\par  
\caption{A typical absorbing configuration of the extended majority-vote model for a lattice of linear size $L=300$. 
Agents with opinion $1$ are painted black and  agents
with opinion $0$ are painted white. There are a total of  $M= 779$ clusters and  the largest one comprises 
$S_m= 4581$ agents. 
\label{fig:3} }
\end{figure}

In this section we first demonstrate that the heterogeneous absorbing configurations  of the extended majority-vote model  do persist in the 
thermodynamic limit and  then  we proceed with the characterization of the opinion domains (clusters)  that fragment those configurations. 
Figure \ref{fig:3} illustrates one such a configuration for $L=300$.
We recall that  a cluster is simply a connected,  bounded lattice region wherein the agents share the same opinion. 
We focus on  a few  relevant statistical quantities to characterize the cluster organization, 
namely,  the average number of  clusters
$\langle M \rangle $, the average size  of the largest
cluster $\langle S_m \rangle$, and the distribution of cluster sizes $P_S$. We must evaluate these quantities
for different lattice sizes  (typically we use $10^4$ samples for each $L$) and then take an appropriate extrapolation procedure 
to infinite lattices ($L \to \infty $).

To simulate efficiently the extended majority-vote model for large lattices we  first make a list of the active agents.  An active agent is an agent 
whose opinion differs from the most frequent opinion among its extended neighborhood. 
Clearly, only active agents can change their opinions and so it is more efficient  to select the
target agent randomly from the list of active agents rather than from the entire lattice.  In the case that
the opinion of the target agent is modified by the majority-vote rule, we 
need to re-examine the active/inactive status of the target agent as well as of all its neighbors so as
to update the list of active agents.  The dynamics is frozen when the list of active agents is empty. The implementation of
a similar  procedure allowed the simulation of Axelrod's model for very large lattices and the clarification of 
several issues regarding the stability of the heterogeneous absorbing configurations of that model \cite{Barbosa_09,Peres_10}.

\subsection{Statistics of the number of clusters}

\begin{figure}[t]
\centerline{\epsfig{width=0.52\textwidth,file=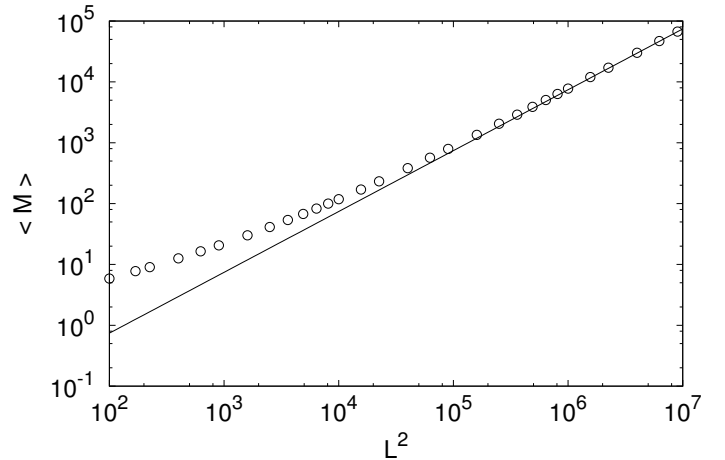}}   
\caption{Logarithmic plot of the average number of clusters $\langle M \rangle $  as function of the
the lattice area $L^2$. The solid straight line is the fitting
$ \langle M \rangle  =  0.00742  L^2 $ for large $L$. Each symbol represents the average over
$10^4$ samples so the sizes of the  error bars  are smaller than the symbol sizes.
\label{fig:4} } 
\end{figure}

We begin our analysis by presenting  the  dependence of the average number of clusters $\langle M \rangle$ 
on the lattice area $L^2$  in Fig.\ \ref{fig:4}. The increase of $\langle M \rangle$
with increasing $L^2$ is the evidence that confirms that the extended majority-vote model
exhibits heterogeneous absorbing configurations in the thermodynamic limit, $L \to \infty$. More to the point, 
we find 
\begin{equation}\label{a_F1}
\lim_{L \to \infty} \frac{\langle M \rangle}{L^2} =  0.00742 \pm 10^{-5} .
\end{equation}
For the purpose of comparison, for frozen random configurations in which the $L^2$ sites are set to $1$ or $0$ 
with the same probability
this ratio yields $0.1342 \pm  0.0001$ (and so the average number of clusters scales linearly with the
lattice area  too). We note that the reciprocal of the ratio given in Eq.\ (\ref{a_F1}) is the average size of the clusters: 
$\langle S \rangle \approx 134.8$ for the extended majority-vote model and $\langle S \rangle \approx 7.45$ for  random configurations.  

\subsection{Statistics of the largest cluster}

\begin{figure}
\centerline{\epsfig{width=0.52\textwidth,file=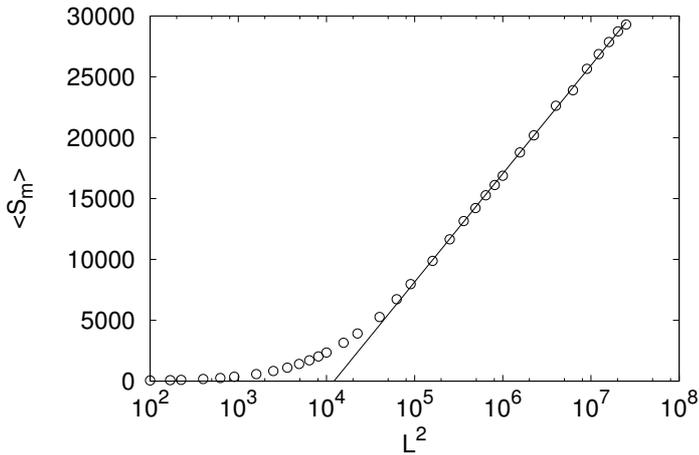}}  
\caption{The average size of the largest cluster $ \langle S_m \rangle$  as function of the
the lattice area $L^2$. The solid line is the fitting $ \langle S_m \rangle = 1933 \ln L^2 - 36363 $. 
Note the logarithmic scale in the $x$-axis.
\label{fig:5} }
\end{figure}

Although the mean size of a typical cluster is finite, there may be clusters
whose sizes can become arbitrarily large as the lattice size increases. To investigate this point we show in
Fig.  \ref{fig:5} the average size of the largest cluster $\langle S_m \rangle$ as function of the lattice area.
As in the case of $\langle M \rangle $, the asymptotic scaling is revealed only for large lattices ($L > 500$) and
it indicates that $\langle S_m \rangle$ increases with $\ln L^2 $ for large  $L$.

In the thermodynamic limit the relevant quantity is then 
\begin{equation}\label{b_F1}
\lim_{L \to \infty} \frac{\langle S_m \rangle}{\ln L^2} =  1933 \pm 10 .
\end{equation}
We found that $\langle S_m \rangle$ scales with $\ln L^2 $ for random configurations also, though the ratio  
$\lim_{L \to \infty} \langle S_m \rangle /\ln L^2 =  53.0 \pm  0.2$ is considerably smaller than the ratio
given in Eq.\ (\ref{b_F1}).

It is interesting to note that the standard deviation $\left [ \left \langle S^2 \right \rangle 
- \left \langle S \right \rangle^2 \right ]^{1/2} $ tends to the constant value   $695.5 \pm 0.5$ in the
thermodynamic  limit. This amounts to a large but finite variance and so in order to satisfy the 
Tchebycheff inequality the set of clusters whose  size grows like $\ln L^2$ must have  measure zero
in that limit. 
Since the  probability that a randomly chosen site belongs to one such a cluster vanishes like 
$ \left ( \ln L \right )/L^2$, the measure of the set of diverging clusters  will be vanishingly small if 
its cardinality is finite, i.e., if only a finite number of clusters have sizes scaling with $ \ln L^2$.

\subsection{Distribution of cluster sizes}

\begin{figure}
\centerline{\epsfig{width=0.52\textwidth,file=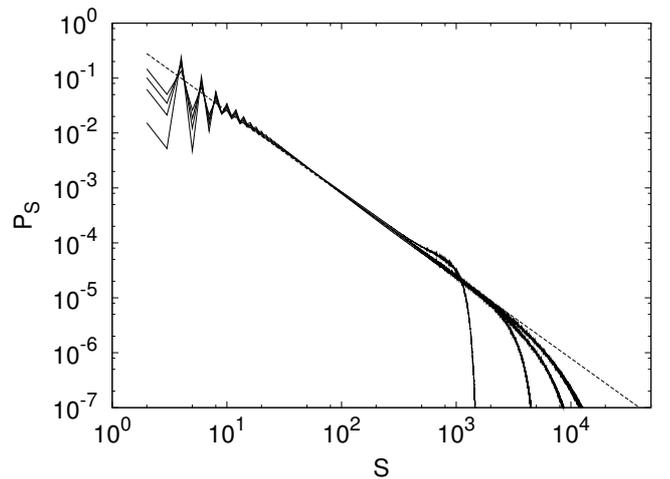}}  
\caption{Logarithmic plot of the distribution of cluster sizes $S$ for  (solid lines from left to right at
$P_S = 10^{-6}$) $L=50, 100, 200$ and
$1000$. The dashed straight line is the fitting $P_s = 0.79 S^{-1.5}$. These distributions are obtained by 
sampling $10^7$ absorbing configurations of the extended majority-vote model.
\label{fig:6} }
\end{figure}

Up to now we found no qualitative differences between the statistical properties of the clusters produced by the extended majority-vote rule or by  
assigning randomly the digits $1$ and $0$ to the sites of the square lattice.  In fact, 
for both types of configurations the  quantities $\langle M \rangle$ and $ \langle S_m \rangle $ exhibit the same scaling behavior 
with the lattice area $L^2$. However, a more detailed view of the cluster organization is given by the distribution $P_S$ of cluster sizes $S$, which is
shown in Fig.\ \ref{fig:6} for different lattice sizes. The data seems to be remarkably well fitted 
by the power-law distribution $P_S \sim S^{-1.5}$ for over more than three decades. 
In addition, the figure indicates that the region where the power-law fitting holds  increases  with increasing $L$. 
This power-law distribution, however, is not compatible with our previous findings of finite 
values for the mean and variance of $P_S$, since the distribution $P_S \sim S^{-1.5}$  has infinite mean and variance. 

\begin{figure}
\centerline{\epsfig{width=0.52\textwidth,file=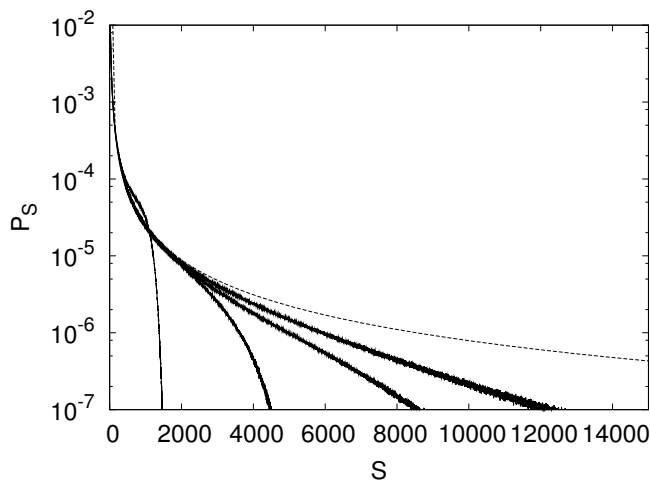}}   
\caption{The same data shown Fig.\ \ref{fig:6} plotted using a semi-logarithmic scale.
\label{fig:7} }
\end{figure}

\begin{figure}
\centerline{\epsfig{width=0.52\textwidth,file=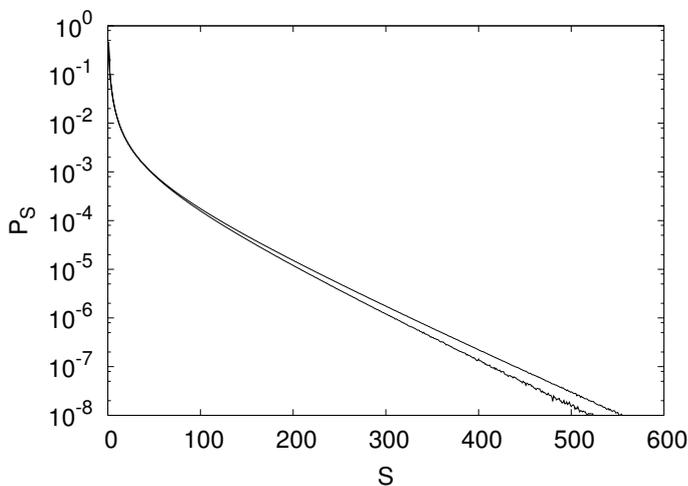}}    
\caption{Semi-logarithmic plot of the distribution of cluster sizes $S$ for random configurations with (solid lines from bottom to top) $L=100$ and
$300$.  These distributions are obtained through
the sampling of $10^8$ random configurations.
\label{fig:8} }
\end{figure}

To answer this conundrum, in Fig.\ \ref{fig:7} we plot the same data using  a semi-logarithmic scale instead of the log-log scale of 
Fig.\ \ref{fig:6}. 
The results show that for large lattice sizes the distribution  $P_S$ is in fact exponential, the power-law behavior 
being valid only in an intermediate range of cluster sizes. The results for the larger lattice $L=1000$ (though not the  results for 
the smaller lattices $L=50$ and $L=100$) show that the exponential distribution is
not a mere cut-off inherent to the finitude of the lattices used in the simulations, as manifested by the straight-line appearance of
$P_S$  revealed by  the semi-log graph in the regime of large clusters. For the sake of comparison, in Fig.\  \ref{fig:8} we show the distribution of cluster sizes for frozen random configurations which can be
well-described by the exponential distribution 
$P_s  \approx 0.00085 \exp \left ( - 0.02 S \right ) $ for  large $S$. The puzzling
power-law behavior found for intermediate values of $S$ in the extended majority-vote model is missing
in this case.

Finally, it is interesting to note that, for any absorbing configuration of the extended majority-vote model, clusters of size $1$ are prohibited, clusters of sizes $2$ and $3$ are allowed at the sides of the square lattice  only, and the only cluster of size $4$ allowed in the bulk of the lattice is that in which the four sites surrounding a unit  cell are in the same state. In fact, this elementary square cluster is  the only structure capable to propagate through the lattice, i.e., whenever a stable cluster is formed one such elementary square is formed.

\subsection{Vulnerability to noise}

\begin{figure}
\centerline{\epsfig{width=0.52\textwidth,file=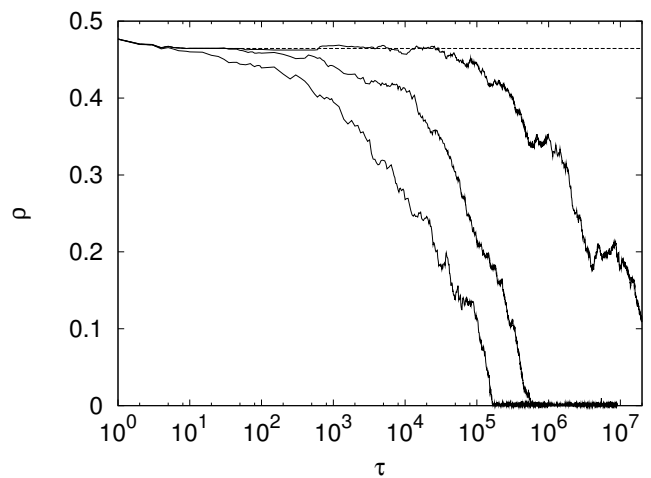}}
\par 
\caption{Fraction of sites in state  $1$ as function of the number of Monte Carlos steps $\tau$ for the  thermal noise  intensity 
(solid lines from left to right)
$p = 10^{-3},10^{-4} $ and $10^{-5}$ and for a lattice of size   $L=50$.   The dashed line is the evolution for the noiseless case which rapidly gets stuck in
an absorbing configuration. The initial configuration is the same for the four trajectories.
\label{fig:9} }
\end{figure}

A word is in order about the robustness of the heterogeneous absorbing configurations to the effect of noise, which allows the
agents to oppose the  opinion of the majority with some small probability. More pointedly, after applying the 
update rule for each agent, which may or may not result in the change of  its opinion, we 
flip  its opinion with probability $p \in \left [ 0,1/2 \right ] $ \cite{deOliveira_92}. 
Since in this case there are no absorbing configurations the notion of active sites is useless and so we 
implement the dynamics by picking lattice sites at random with the same probability.   Then the unit of  
time (a Monte Carlo step) corresponds to the update of 
$L^2$ randomly chosen sites. For $L =50$ we present in  Fig.\ \ref{fig:9} the time evolution of 
the fraction of sites in state $1$, which we denote by $\rho$ as in Sect. \ref{sec:MF}, 
for different values of the 
noise parameter $p$ but for  the same initial configuration.  This figure, which exhibits the evolution of 
a single sample, shows the vulnerability of the heterogeneous absorbing configurations against a vanishingly amount of thermal  
noise if one gives enough time  for thermalization.  For the small values of $p$ considered, the  steady-state is  close to 
one of the two homogeneous configurations, with all individuals exhibiting  opinion $0$ or opinion $1$. The specific outcome is a stochastic 
process which  depends on the choice of the initial configuration as well as on  the sequence of
random numbers  used in the update rule.

\section{Conclusion}\label{sec:conc}

This paper is motivated by the claim that the extended majority-vote model, for which the target site counts in the reckoning of the  majority,  exhibits a nontrivial multicultural regime, i.e., heterogeneous absorbing configurations \cite{Parisi_03}. From the perspective of the statistical physics, we find this claim
 most intriguing, as  we expect the model to behave similarly to the 
majority-vote models commonly considered in the physics literature, which are basically Ising models with
zero-temperature Glauber kinetics \cite{Glauber_63}, and so exhibit only two  homogeneous absorbing  configurations in the thermodynamic limit.  Our first impression was that the heterogeneous absorbing configurations were
artifacts of the small lattice size ($L=20$)  used or of some  `non-physical' elements of the
original model such as the nearest and next-nearest neighbors interactions (Moore neighborhood) and the
parallel update of the sites  \cite{Parisi_03}.

Accordingly, we have modified the original model proposed by Parisi et al.  by introducing the usual 
nearest neighbors interactions (von Neumann neighborhood) and the random sequential update of sites. A
careful finite size scaling analysis of a few relevant measures that characterize the absorbing configurations shows
that the heterogeneous  absorbing configurations not only  persist in the thermodynamic limit but 
produce a tricky distribution of cluster sizes $P_S$ that decays as a power law $P_S \sim S^{-1.5}$ for clusters of  intermediate  size $S$ and as an exponential for very large clusters sizes (see Figs.\ \ref{fig:6} and
\ref{fig:7}). Essentially, the reason for this somewhat unexpected outcome is the criterion of update in case of a tie for  a site in the bulk 
of the lattice: 
the site remains unchanged 
rather than flipping to another state with probability $1/2$ as usually done in statistical mechanics models \cite{Tome_91,deOliveira_92}. 
What is remarkable is that the existence of these heterogeneous absorbing configurations for the extended majority-vote model, 
as well as their absence in the  more usual variants of that model, can actually be predicted analytically using the pair approximation: the multiple-clusters regime is signaled by the presence of an infinity of attractive fixed points of the mean-field equations.

We found that the statistical properties (e.g., average number of clusters, mean size of the largest cluster and the asymptotic distribution of cluster sizes)  of the clusters that break up the absorbing configurations (see Fig.\ \ref{fig:3}) are qualitatively identical to those of random configurations. This could also  be inferred by the very short convergence times  (see dashed line in Fig.\ \ref{fig:9}) which indicate that the absorbing configurations are  very close to  the initial random configurations, since a few lattice updates suffice to reach them. We are not aware of any similar study of the statistical properties of   the heterogeneous configurations of Axelrod's model, so we cannot  compare the cluster organization of these two models.

Our findings, which corroborate the results of   Parisi et al.  \cite{Parisi_03}, shows that
the extended majority-vote model does exhibit  a  `multicultural' regime, despite the fact that the update rule
biases the agents  to become more similar  to each other.  A similar conclusion holds for Axelrod's model 
as well \cite{Axelrod_97,Castellano_00,Barbosa_09}, except that in Axelrod's model the similarity is a prerequisite for interactions - the  `birds of a feather flock together'
hypothesis which states that individuals who are similar to each other are more likely to interact
and then become even more similar \cite{Moscovici_85}. (A similar assumption has been used to model the
interspecies interactions in spin-glass like model ecosystem \cite{deOliviera_02}.) 
The majority-vote model is considerably simpler and converges to the absorbing configurations much faster
than Axelrod's. However, the (short) inventory of advantages stops here:  in disagreement with the claim of Parisi et al.  \cite{Parisi_03} we found
that the absorbing configurations of the extended majority-vote model are  
vulnerable to the noisy effect of flipping the opinions of the agents with some small probability.
It is well-known that this type of noise destabilizes the heterogeneous configurations of Axelrod's model too \cite{Klemm_03c}. 
Of course, the extended majority-vote model lacks the main appealing feature of Axelrod's model, namely, 
the existence of a  non-equilibrium phase transition 
that separates the homogeneous and the polarized regimes in the thermodynamic limit. In that sense it would be
interesting to find out how a similar transition could be induced in the zero-temperature majority-vote model.

\begin{acknowledgments}
This research   was  supported by  The Southern Office of Aerospace Research and Development (SOARD), grant FA9550-10-1-0006,  
and Conselho Nacional de Desenvolvimento Cient\'{\i}fico e Tecnol\'ogico (CNPq).
\end{acknowledgments}



\begin{thebibliography}{99}

\bibitem{Latane_81} B. Latan\'e, {\it American Psychologist} {\bf 36}, 343 (1981).

\bibitem{Moscovici_85} S. Moscovici, {\it Handbook of Social Psychology} {\bf 2}, 347 (1985).

\bibitem{Boyd_85} R. Boyd and P. J. Richerson, {\it Culture and the evolutionary process} 
(University of Chicago Press, Chicago, 1985).

\bibitem{Nowak_90} A. Nowak, J. Szamrej and B. Latan\'{e}, {\it Psychol. Rev.} {\bf 97}, 362 (1990).

\bibitem{RMP} C. Castellano, S. Fortunato and V. Loreto, {\it Rev. Mod. Phys.} {\bf 81}, 591 (2009).

\bibitem{Axelrod_97} R. Axelrod, {\it J. Conflict Res.} {\bf 41}, 203 (1997).

\bibitem{Kennedy_98} J. Kennedy, {\it J. Conflict Res.} {\bf 42}, 56 (1998).

\bibitem{Klemm_03c} K. Klemm, V. M. Egu\'{\i}luz, R. Toral, M. San Miguel, 
{\it Phys. Rev. E} {\bf  67}, 045101R (2003).

\bibitem{Greig_02} J. M. Greig, {\it  J. Conflict Res.} {\bf 46}, 225 (2002).

\bibitem{Klemm_03b} K. Klemm, V. M. Egu\'{\i}luz, R. Toral, M. San Miguel,  {\it Physica A} {\bf 327}, 1 (2003).

\bibitem{Klemm_03a} K. Klemm, V. M. Egu\'{\i}luz, R. Toral, M. San Miguel, {\it Phys. Rev. E} {\bf 67}, 026120 (2003).

\bibitem{Marro_99} J. Marro and R. Dickman, {\it Nonequilibrium Phase Transitions in 
Lattice Models} (Cambridge University Press, Cambridge, UK, 1999).

\bibitem{Castellano_00} C. Castellano, M. Marsili and A. Vespignani,  
{\it Phys. Rev. Lett.} {\bf 85}, 3536 (2000).

\bibitem{Barbosa_09} L. A. Barbosa and J. F. Fontanari, {\it Theory Biosci.} {\bf 128}, 205 (2009).

\bibitem{Parisi_03} D. Parisi, F. Cecconi,  F. Natale, {\it J. Conflict Res.} {\bf 47}, 163 (2003).

\bibitem{Liggett_85} T. M. Liggett, {\it Interacting Particle Systems} (Springer, New York, 1985).

\bibitem{Gray_85} L. Gray, in  {\it Particle systems, random media, and large deviations} (American Mathematical Society, Providence, RI, 1985), Vol. 41, pp. 149--160.

\bibitem{Tome_91} T. Tom\'e, M.J.  de Oliveira and M.A.  Santos,  {\it J. Phys. A: Math. Gen.} {\bf 24} 3677  (1991).

\bibitem{deOliveira_92} M. J. de Oliveira, {\it J. Stat. Phys.} {\bf 66}, 273 (1992).


\bibitem{Glauber_63} R. J. Glauber, {\it J. Math. Phys.} {\bf 4}, 294 (1963).

\bibitem{Peres_10} L. R. Peres and J. F. Fontanari,  {\it J. Phys. A: Math. Theor.} {\bf 43},  055003 (2010).



\bibitem{deOliviera_02} V. M. de Oliveira and J. F. Fontanari,  {\it Phys. Rev. Lett.}  {\bf 89}, 148101 (2002).









\end{thebibliography}
\end{document}